# Are All Probabilities Fundamentally Quantum Mechanical?

Rajat Kumar Pradhan*
Rajendra College, Bolangir, Orissa, India-767002.

(Date: 10.05.2011)

**Abstract**

The subjective and the objective aspects of probabilities are incorporated in a simple duality axiom inspired by observer participation in quantum theory. Transcending the classical notion of probabilities, it is proposed and demonstrated that all probabilities may be fundamentally quantum mechanical in the sense that they may all be derived from the corresponding amplitudes. The classical coin-toss and the quantum double slit interference experiments are discussed as illustrative prototype examples. Absence of multi-order quantum interference effects in multiple-slit experiments and the Experimental tests of complementarity in Wheeler's delayed-choice type experiments are explained using the involvement of the observer.

*email: rajat@iopb.res.in

## 1. Introduction

It is well known that probabilities have a dual character- they may be either subjective or objective or may be partly subjective and partly objective [1-6]. Accordingly, the corresponding interpretations are called subjective or belief interpretation, objective or frequency interpretation and combined or support interpretation respectively. The Bayesian approach, in particular, is an attempt to incorporate the subjective probability assignments alongside the objective probabilities to give probability theory its much-needed completeness. In fact, there have been attempts [7, 8, 9] to interpret the probabilities of quantum theory by appealing to Bayesian methods of inference, but never the amplitudes themselves. The reason is that the amplitudes are inherently complex quantities violating Kolomogorov's first axiom and hence do not qualify to be called probabilities. Therefore, a satisfactory understanding of the Born rule from more fundamental principles is lacking. On the other hand, the phenomenal success of quantum theory seems to call for a reappraisal of our primitive classical notion of probabilities themselves, to which end proposals like negative (extended) probabilities[10-16] and complex (exotic) probabilities [17, 18] have been put forward in the literature to make sense of quantum theory.

The dual nature of probabilities is a consequence of the inevitable fact that *'the best that one can do in the case of an unobserved event is to assign probabilities in regard to the possible outcomes in keeping with the information that one has about it'*. This fundamental fact is utilised to propose a **duality axiom** for probabilities which incorporates both aspects viz. the objective facts like frequencies or weights, and the possession (or lack) of subjective knowledge of such facts. The subjectivity, as far as it is to be scientifically meaningful, can enter only as the vehicle of assignment of probabilities, since a probability is not something out there in the objective world- the world that is made up of objectively observed and observable facts only- but it is we, the subjects, who make probability assignments for unobserved and unobservable events (past, present and future) on the basis of the known facts and thus the probabilities serve the purpose of being a measure of our ignorance as well as of our knowledge. To be more explicit, the biasedness of a coin, loadedness of a die and the frequency of outcome etc. are objective, while the probability assigned to a single outcome is subjective based on the knowledge of such objective facts. This is precisely what we term here as the dual nature of probability.

Gaiffman [19] rightly points out that the most rational probability assignment must be on the basis of the most complete information about the event. This, obviously, is the 'expert assignment' [20] which is to be adopted by all other subjects in order not to make off-the-track assignments. Now, science can be done only with the most complete information about the system or event and thus the expert probability is the most suitable one for scientific investigations and we shall assume in this work that the subjects are experts who in their subjective assignments make use of the maximum information available to anyone anywhere upto that instant. This is necessitated by the fact that we need to have testable predictions in science, and not just arbitrary personal guesses based on inadequate knowledge of facts about the system or event. It turns out that with this kind of necessary restrictions on the subjective probability, we can tackle the dual nature of probabilities to a large extent using the extremely simple and most general kind of formulation proposed here.

It will be shown that once we try to incorporate both, the subjective and the objective probabilities, we must have fundamentally complex assignments in order not to violate the Kolomogorov axioms. Thus, the link with quantum mechanics is very easily established and it also paves the way for a fresh interpretation of quantum mechanics itself based on subject-object participation as envisaged by Wheeler [21] and conforming to von Neumann's original line of reasoning [22]. Recently, Manousakis [23] has attempted to base quantum theory on the basis of operations of consciousness and Stapp [24] has investigated the mind-brain connection to pinpoint the role of consciousness in the collapse of a quantum state.

In section-2 we propose to tackle the duality of probabilities in the form of an axiom. We motivate the need for this duality axiom by a simple analysis of subjectivity and objectivity in science. We impose, following common sense, a **sanity requirement** which requires the assigning subject to reflect the objective facts about the event in its subjective assignment. In section-3 we discuss the classical prototype in the form of the coin toss and in section-4 the quantum prototype in the form of the double slit interference. In section-5 we draw attention to the absence of higher order interference effects in experiments and show that the formulation proposed here does indeed very naturally explain such absence. Similarly, the vanishing of quantum interference in Wheeler's delayed choice type experiments are also easily explained. We conclude in the end with a discussion of the advantages, caveats, and the future directions for further development.

## 2. The Duality Axiom for Probabilities

To motivate the introduction of the Duality Axiom, let's first discuss subject-object duality. For all purposes, we may define as objective that which is common to all subjects at any particular instant of time. As an example, Gravitation existed before it became an objective fact primarily through the subjective efforts of Newton, while tachyons or magnetic monopoles exist only subjectively (as of now), still waiting to be granted objective existence by experimental proof. Please mark how the objective existence depends on the subjective assent for its so-called objectivity and, till that moment how it is regarded as mere fiction and not fact, no matter how compelling the theoretical arguments for its existence may be.

An event at any instant fails to be objective if there are subjective differences regarding it at that instant. In fact, the so called objective world independent of any perceiving subject is but an extrapolation on which all subjects agree, thus granting it its objectivity. Even if only one subject does not agree, the issue fails to be objective. The objective should obviously have the concurrence of every subject. Similarly, the case of 'no evidence' means 'no objective evidence' and 'symmetric evidence' means 'symmetric objective evidence'. Even if just one subject has access to some fact about an event, then the ignorance of that fact loses its objectivity. Not only should knowledge be objective but ignorance should also be objective, and this has to be taken care of while dealing with the dual aspects of probabilities. If ignorance is not objective i.e. common to all subjects, then probabilities will lose their objectivity and subjective predictions will be widely (and sometimes wildly) different making them scientifically meaningless.

Further, any probability assigned on the basis of insufficient objective knowledge, including knowledge of what any other subject might have known about the event, is not a dependable assignment and thus falls out of scientific reckoning, unless and until it is modified on the basis of more complete information. If nobody knows anything about an event, then only we have 'no evidence' about that event. If even one person has some information about the event, then that information is to be made objective by way of subject-to-subject information transfer or by way of repetition of the process of attaining the same, for otherwise subjective assignments will vary arbitrarily because of their being made on the basis of unequal information. In the absence of any objective evidence whatsoever, the subjective assignment is automatically guided by the 'symmetric evidence' formula since that is the best that we can do subject to subsequent verification. In that case, the subjective assignment of assumed *a priori* equality of the probability of all outcomes determines the objective amplitude.

Let's now examine how we actually assign a probability. Clearly, we need the following five facts for the purpose:

- Knowledge of the particular outcome e, for which the probability is to be assigned
- Knowledge of the full sample space of all possible outcomes Ω
- Lack of knowledge of the outcome in case the event has already occurred, otherwise the probability assignment becomes trivial i.e. P(e) = 1 or 0, depending on whether it has already occurred or not.
- Knowledge of the statistical frequency f of occurrence of the particular outcome e in question
- In the absence any knowledge whatsoever of the frequencies, allotment of equal *a priori* frequency to all outcomes

This, in itself, is sufficient evidence for the involvement of the knowing subject (alongside the objective facts) through the knowledge or lack of it in each step as enlisted above in the assignment of a probability, since any knowledge or lack of it is but a matter of stored data in the memory of the assigning subject consequent upon the interaction (mostly sensory) of the knowing subject with the object in question. The objective sample space, the particular objective outcome selected from it, the objective statistical frequency f of the outcome etc. all indeed have their subjective counterparts or images in the memory storage, which is a *sine qua non* for the purpose for any perceptual knowledge by comparison/matching and, failing which there can never arise any question of assigning a probability. This argument firmly establishes that the dual nature of probabilities is not a conjecture but a fact in need of a satisfactory formulation.

However, the fact that the subjective probability for an event varies from person to person renders any objective study of such assessments meaningless and this also acts as a deterrent for any systematic investigation of such arbitrary personal probabilities. For example, the subjective probability for a so-called improbable event (objective classical probability 0) may even be assigned a value 1 (subjective certainty) because of subjective ignorance of the said improbability. In the face of such absurdities, we have to devise ways to safeguard the veracity of the objective classical probability based on objective facts while at the same time incorporating rational subjective assessments.

To this end, we formulate the following **Duality Axiom** for probability:

**Duality Axiom**: *The probability P(e) of an event E in a probability space (Ω, E, P) with sample space Ω, event space E and Probability measure P is a function of both, the objective amplitude $A_o$ and the subjective assignment $A_s$.*

$$P(e) = \xi(A_s, A_o) \qquad \ldots \ldots \ldots \ldots \quad (1)$$

The subjective and the objective assignments are, in general, independent of each other in which case we may write the expression in the simplest possible case as the trivial bilinear:

$$P(e) = A_s \cdot A_o \qquad \ldots \ldots \ldots \ldots \quad (2)$$

and, they may be complex numbers as required by the Kolomogorov axioms for the probabilities P(e). But, for reasons discussed above we shall have to demand that *the subjective assignment somehow should reflect the objective amplitude*. This is formulated as the *sanity requirement* (see below). All facts about the system are encoded in the objective amplitude $A_o$ which may be calculated from the frequency of the event in a large number of identical trials or by solving an equation like the Schrodinger equation as the case may demand, while $A_s$ is simply the expression of the subjective knowledge of the same objective facts.

We note that what is known from classical probability theory is only the quantity on the left i.e. the probability P(e), which is the objectively ascertained (or ascertainable) probability based on frequencies or statistics and which satisfies the three well-known Kolomogorov axioms. As far as classical probabilities are concerned, Kolomogorov's axioms suffice, but when we enter the quantum arena, our traditional notion of probability fails. Therefore, the duality axiom will be useful not only to determine P(e) in terms of subjective and objective assignments but also to tackle the interpretational problems of both Probability theory and of Quantum theory simultaneously.

We shall now determine some of the properties of the assignments by appealing to Kolomogorov's axioms. First of all, for simplicity and for the sake of logical consistency, we state the sanity requirement:

- *sanity requirement* : The subjective assignment must reflect the objective amplitude.

This serves to keep out wild and logically untenable subjective assignments made without having proper objective information about the system or event in question. Further, here the objective amplitude and the subjective assignment both have the identical information content. This, of course, makes the conscious subject a somewhat passive assigner, stripped of any free will. But, this is a small price to pay in order to safeguard scientific rationality, while at the same time including the conscious observer in the scheme.

It is easy to see that with real assignments $A_s$, $A_o \in \mathbf{R}$, it is never possible to satisfy all the Kolomogorov axioms. Thus they have to be complex numbers and the sanity requirement then fixes the dependence of $A_s$ on $A_o$ as:

$$A_s = A_o^* \qquad \ldots \ldots \ldots \ldots (3)$$

Further, appealing to Kolomogorov's second axiom about the unit probability measure on the measure space $(\Omega, E, P)$ i.e. $P(\Omega) = 1$ we may absorb a possible multiplicative constant as a normalization factor in the assignments without any loss of generality and write finally :

$$P(e) = A_s \cdot A_o = A_o^* \cdot A_o = A^2 \qquad \ldots \ldots \ldots \ldots (4)$$

where, $A_o = A_s^* = A \exp(i\delta)$, $A = |A_s| = |A_o| \geq 0$ is their common magnitude and $\delta$ is an arbitrary phase.

To make the correspondence of the probability with the frequency, we may write the objective amplitude explicitly as:

$$A_o = \lim_{N \to \infty} \{\sqrt{(f/N)}\} \exp(i\delta) \qquad \ldots \ldots \ldots (5)$$

where f is the frequency of the outcome in a very large number of identical trials N.

In classical probability theory, the phase factor will be used only as a guarantor of consistency. For example, if we have mutually independent events 1 and 2 with amplitudes $A_1$ and $A_2$, we can write the total objective amplitude as $A_o = A_1 \cdot A_2$ such that the probability of both events together becomes $P_{12} = A_s \cdot A_o = |A_o|^2 = |A_1|^2 \cdot |A_2|^2 = P_1 \cdot P_2$, as expected classically.

In quantum mechanical applications, the phase factor will play its usual role as the action in Planck units (i.e. $\delta = S/\hbar$) as the correspondence then will be established not through eq. (5), but by demanding that the amplitude $A_o$ be the wave function i.e. a solution of the Schrodinger equation exactly as in the path integral approach to quantum mechanics.

We shall now apply the formulation to study the prototype classical and quantum probabilities which will be sufficient to convince us that the formulation can be applied to all other problems without any difficulty.

## 3. Application to classical probability- coin toss

Let's consider a single toss of the classical coin as a prototype of classical probability. If one knows about the unbiasedness then, on employing $A_s^* = A_o$ and $A_o(h\cup t) = A_o(h) + A_o(t) = |A_o(h)|\exp(i\delta_h) + |A_o(t)|\exp(i\delta_t)$, the total probability of obtaining a head(h) or a tail(t) becomes:

$$P(\Omega) = P(h\cup t) = |A_o(h\cup t)|^2 = |A_o(h)+A_o(t)|^2 = P(h) + P(t) + 2|A_o(h)||A_o(t)|\cos(\delta_h - \delta_t) \quad \ldots (6)$$

It is obvious that the mutually exclusive outcomes of an event may, in general, be taken to represent the orthogonal basis vectors (here |h> and |t>) of a Hilbert space spanned by them with the amplitudes providing the coefficients just as in quantum mechanics. Classically, however, the interference term vanishes because of the fact that the two states |h> and |t> have no possibility of simultaneous occurrence i.e. $P(h\cap t) = 0$. This 'classical orthogonality' (i.e. mutual exclusivity or disjointedness) of the final states makes any quantum-like superposition impossible and meaningless for a thin coin and fixes the value of the relative phase i.e. $\cos(\delta_h - \delta_t) = \cos(\pi/2) = 0$ for all times, so that we have the familiar classical result reproduced:

$$P(h\cup t) = P(h) + P(t) = 1 \qquad \ldots \ldots \ldots (7)$$

For a thick coin discussed in detail by Yong and Mahadevan [25], however, we would have a finite probability for it to land on the sides so that eq.(4) is to be replaced by $P(h\cup t\cup s) = P(h) + P(t) + P(s) = 1$, with the condition that the outcomes are pair-wise disjoint i.e. $\delta_h - \delta_t = \delta_t - \delta_s = \delta_s - \delta_h = \pi/2$.

In general, a random classical event with n possible outcomes can be represented by an n-dimensional Hilbert space with the objective assignments $A_o(i)$ along the n orthogonal coordinates such that the relative phases satisfy the time-independent pair-wise disjoint condition viz. $\cos(\delta_i - \delta_j) = \cos(\pi/2) = 0$ always, for all $i \neq j$, such that $P(\Omega) = \Sigma_j P(j) = 1$ is satisfied. In this kind of Hilbert space-like description of the classical event, the subjective assignments $A_s(i) = A_o(i)^*$ span the dual Hilbert space.

Classically, as far as the probabilities of the outcomes are concerned, superposition of the assignments fails to be effective in producing any interference effects because of the fact that each classical outcome entails necessarily a measurement which destroys all interference by projecting it along any one of the coordinates(basis vectors). It is only in Quantum theory that the relative phases have observable consequences because of quantum interference as in the famous double slit experiment. Nevertheless, we have succeeded in reformulating probability theory taking into account the subjectivity of probabilities as an equally important factor alongside the objectivity in determining the probability.

Now, as a precursor to the application to quantum theory for a final vindication of the dual nature of probabilities we shall briefly discuss further the coin toss in regard to possible quantum effects in light of the present approach.

When the coin is tossed, it has two possibilities of landing and the assigner has also the images of these two possibilities in his mind when he tries to call. We assume the objective and the subjective possibilities to be unbiased and independent of each other. There are four possible outcomes in all, namely, h*h, h*t, t*h and t*t, where the first one is the subjective assignment and the second one is the objective outcome. The probability of a correct guess is clearly ½, which is also the probability of a wrong guess. The subjective 'call' will tally with the objective 'fall' only half the time and this is how the probability is actually and operationally assigned by the caller to the outcomes h or t!

Only after the caller physically comes to know the state (h or t) after the toss, does the doubt in the mind vanish, giving rise to the observed state and not before that. Thus, *"It is the subjective doubt that we have been objectively describing as a superposition in quantum mechanics, and, as already mentioned earlier, this objectivity surely results from the unanimity about the same doubt shared by all subjects involved"*. This is the von Neumann subjective state collapse, which again gets an objective status when the same knowledge of the state of the coin (h or t) becomes common to all subjects involved. *The so-called observer-independence of the classical world of objective science is thus only an assumption and not a final fact and is certainly not true in quantum mechanics.*

## 4. Application to quantum probability- double slit interference

In the traditional as well as other proposed interpretations, the double slit experiment acquires a central place, both, for its simplicity and profundity. While the interference effects for multiple quanta are explicable parallel to the classical wave mechanical treatment, the quantum interference observed- even when only a single quantum is allowed at a time to transit from the source to the screen- is certainly not explicable without recourse to the possibility of its simultaneous transit through both slits contrary to its particle nature.

Let's now consider the double slit experiment with single quantum in light of the duality axiom and see its interpretational advantages. Assume that a source 's' emits the quanta which travel through the slits 1 and 2 and interfere at position y on the screen. The objective assignment for the process will be the sum of the two mutually exclusive amplitudes $A_o = A_1 + A_2$ and consequently, by the sanity requirement the subjective assignment is $A_s = A_o{}^*$. The probability of arrival at y is then

$$P(y) = A_s \cdot A_o = |A_o|^2 = |A_1|^2 + |A_2|^2 + A_1{}^*A_2 + A_2{}^*A_1 \qquad \ldots \ldots \ldots (8)$$

For each quantum if we write the objective assignments $A_{1,2}$ for passage through slits 1(2) as a product of the independent amplitudes to go from s to 1(2) and then from 1(2) to y, then the arrival probability P(y) becomes:

$$P(y) = |<y|1><1|s> + <y|2><2|s>|^2 = <s|1><1|y><y|1><1|s> + <s|2><2|y><y|2><2|s>$$
$$+ <s|1><1|y><y|2><2|s> + <s|2><2|y><y|1><1|s> \quad \ldots (9)$$

It is seen that each of the four terms is a product of the subjective and the objective assignments (here amplitudes) in keeping with the duality axiom. But, the noteworthy feature is that ``***the interference arises not from an actual passage of the single quantum through both slits simultaneously but through the interplay of the subjective assignment for its passage through one slit with the objective amplitude for passage through the other***" as evidenced by the structure of the last two terms. In the interference region the actual or objective passage through slit 1 is taken in conjunction with the assumed(subjective) assignment (due to lack of which-slit knowledge) for passage through slit 2 and vice versa. These two interference terms combined together as two mutually exclusive ways of realising the interference at the point y give a real positive probability:

$$P(1 \cap 2) = A_1^* A_2 + A_2^* A_1 = 2|A_1||A_2|\cos \delta, \ (\delta = \delta_2 - \delta_1 = \text{phase difference}) \quad \ldots (10)$$

This follows from the application of the Kolomogorov axiom regarding $P(A \cup B)$ and comparing it with eq.(6) as follows:

$$P(y) = P(1 \cup 2) = P(1) + P(2) - P(1 \cap 2)$$
$$= [P(1 \text{ only}) + P(1 \cap 2)] + [P(2 \text{ only}) + P(1 \cap 2)] - P(1 \cap 2)$$

or,
$$P(1 \cup 2) = P(1 \text{ only}) + P(2 \text{ only}) + P(1 \cap 2) \quad \ldots \ldots \ldots (11)$$

This is to be contrasted with the non-classical total probability formula proposed recently by Khrennikov [26] on the basis of stochastic modelling. Also note that P(1 only) and P(2 only) are the respective probabilities for passage through slits 1 only & 2 only when both slits are open:

$$P(1,2 \text{ only}) = P(1,2)|_{\text{both open}} = P(1,2)|_{2,1 \text{ closed}} - P(1 \cap 2) \quad \ldots \ldots \ldots (12)$$

Classically, because of mutual exclusivity, the quantity P(1) should not depend on whether we have slit 2 open or closed. But, quantum mechanically we have eq. (9) which tells us that when both slits are open we have no way to tell which path exactly was taken by the quantum. This is because the exclusivity is now at the level of the more fundamental amplitudes rather than at the level of probabilities. The peculiarity of the quantum interference effects is that while the final probability and the single path probabilities are always non-negative, the interference probability $P(1 \cap 2)$ given by eq. (10) may be positive, negative or zero depending on the relative phase $\delta$.

Now, as discussed in the coin toss example above in section-3, it again turns out that ***the probabilities reside as much in the subjective lack of knowledge as in the objective event or the outcome***.

In the interpretation proposed here, we may assign path 1 when it has arrived through 1, path 2 when it has arrived through 2, path 1 when it has arrived through 2 and path 2 when it has arrived through 1. These four possibilities constitute the four terms in eq. (9) above which are simple and self-explanatory as depicted in the following figures in the order in which they appear.

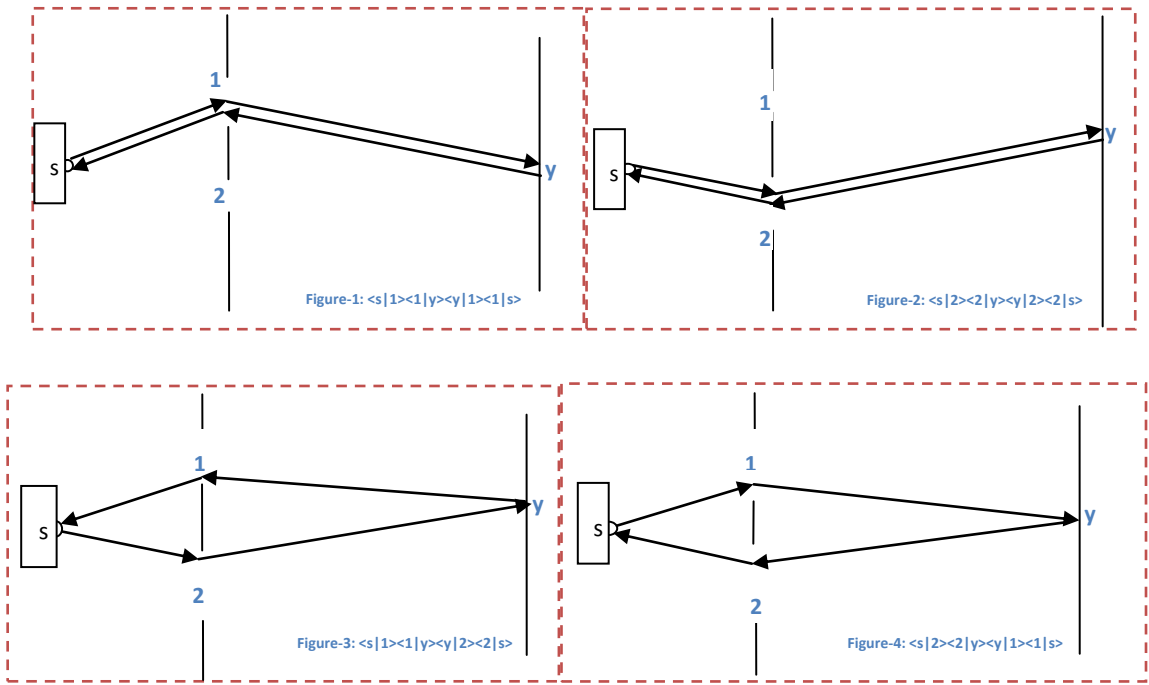

Figure-1: <s|1><1|y><y|1><1|s>

Figure-2: <s|2><2|y><y|2><2|s>

Figure-3: <s|1><1|y><y|2><2|s>

Figure-4: <s|2><2|y><y|1><1|s>

Please note that Soucek[18] has also proposed similar interpretative diagrams treating the amplitudes themselves as complex probabilities, while in Cramer's transactional interpretation [27] the backward arrows from arrival point y to the source s represent the advanced echoes which confirm the transaction in each case.

## 5. Absence of multi-order interference and the delayed-choice experiment

We now discuss the important issue of the absence of higher order interference effects [28] i.e. interference due to simultaneous passage of the quantum through three or more slits. If we assume, as is traditionally done, that the double-slit interference occurs due to simultaneous passage of the quantum through both slits, there is no reason why there should not be higher order interference effects due to similar simultaneous passage through three or more slits. In fact, between slits 1 and 2 shown in the above figures, we can always cut more slits and such higher interference must be expected to occur quite naturally. However, this is not supported by experiments as shown by Sinha *et al* [29], while there is nothing in quantum theory itself or in any of its interpretations proposed so far, which would rule out such third and higher order interference effects.

In our formalism, for an N-slit interference, the arrival probability will be given by only sum over pair-wise terms including self pairs (i=j):

$$P(y) = \Sigma_{i,j} A_i^* A_j = \Sigma_{i,j} (\langle s|i\rangle \langle i|y\rangle)(\langle y|j\rangle \langle j|s\rangle) \qquad \ldots \ldots \ldots (13)$$

where i and j are the slit indices running from 1 to N corresponding to the subjectively assigned and the objectively traversed paths respectively. Since the common sense observation that a quantum can pass only through only one slit at a time is restored in this formulation we automatically come to the conclusion that the subjective doubt also will be about its passage through only one slit at a time. This leads to only two-slit interference terms as in eq. (13) above.

Similarly, another example which is directly related to the role of the observer in quantum theory is the celebrated delayed choice experiment proposed by Wheeler [21] and experimentally tested by Jacques *et al* [30] as proof of the complementarity principle of Bohr that the wave or particle behavior of a quantum depends on the type of experiment the observer chooses to perform. The idea is to get the which-slit information long after the quantum has passed the slit region, and possibly interfered with itself in the screen region, before being detected by either or both of a pair of detectors $D_1$ and $D_2$ placed facing pointedly at slits 1 and 2 respectively such that if one of them clicks it means that the quantum passed through the corresponding slit and not through the other. It turned out that the wave nature (and hence the interference pattern) vanished when the which-path information (and hence particle nature) was ascertained by detection.

Now, in our dualistic formulation, when the arrival point y is replaced by detector $D_1$ or $D_2$, only one of the first two terms in eq.(9) contributes. The interference terms vanish when such detectors are placed as would allow incoming quanta only from the corresponding slit. This is because in such a case only one of the two terms, $P_1 = (\langle s|1\rangle \langle 1|D_1\rangle)(\langle D_1|1\rangle \langle 1|s\rangle)$ and $P_2 = (\langle s|2\rangle \langle 2|D_2\rangle)(\langle D_2|2\rangle \langle 2|s\rangle)$, is realised and it explains the result quite simply. The knowledge of the observer (from the very way in which the experiment is designed by her) leads to the fact of her discarding interference terms and to the fact that there will be no interference and that only one of the detectors will click. Thus, when the screen is replaced by the detectors, it is not a surprise that interference (wave nature) is not observed in delayed-choice type experiments, since it is expected not to occur.

Note however that in our scheme, when the detectors are in place, the total probability P is the sum of the disjoint probabilities $P_1$ and $P_2$, and unless and until the detectors are examined as to which one clicked, the which-path information is not obtained although the quantum interference is lost. Then the situation reduces to the classical coin-toss as discussed in section-3.

## 6. Discussion and conclusion

The main idea is that the dual nature of probabilities is a fact which can no longer be avoided or postponed. Rather, it is the primary issue to be tackled if we wish to remove quantum weirdness and restore common sense in quantum theory. That quantum mechanics could be interpreted probabilistically means that probabilities also share, at least to some extent, the same uncommon features with the former [31]. A little reflection reveals that one of these commonalities has to be the involvement of the conscious observer, though its incorporation is rather difficult and has found stiff opposition in both the fields. More than anything else, this is primarily because of our inability to comprehend and mathematically represent consciousness and its operations in a rational scientific manner. However, as shown in this work we can very simply incorporate the conscious observer into the probability theory through the duality axiom. Simplicity (duality axiom) and common sense (sanity requirement) have been the guiding principles in the present work and, we have successfully implemented the scheme in case of both classical as well as quantum probabilities.

It is demonstrated that in place of the familiar classical probability P(e), we may as well work with the more fundamental assignments which obviously hold up a promise not only to unify all the interpretations (belief, frequency and support interpretations) of probability via the Duality axiom but also for a fundamental new interpretation of Quantum mechanics treating the observer(subject) on par with the observed(object) as envisaged by Wheeler in his participatory universe scenario. The interpretation along with its implications for various curious quantum phenomena like wave-particle duality, delayed choice experiments and non-locality etc. will be treated in a forthcoming paper by the author [32].

Several caveats in the formulation are also in order for being addressed. First of all, the addition of mutually exclusive amplitudes is taken straightaway from quantum theory without sufficient reason. 'Why should the mutually exclusive quantum amplitudes add just like mutually exclusive classical probabilities, if not for matching with experimental findings?'- is the moot question still persisting. The only way out seems to appeal to common sense to justify the procedure that when a number of mutually exclusive alternatives are available for an outcome, the total amplitude for it is the sum of the amplitudes for individual alternatives *a la* classical probabilities. The classical exclusivity is lost as we go beneath the classical probabilities and it appears now at the level of the amplitudes for the alternatives (paths in the double slit example), and therefore, we add the amplitudes for them to get the total amplitude and then move on to find the probability by multiplying with the subjective assignment which is the complex conjugate of the total amplitude. Further, while the amplitudes are mutually exclusive the probabilities are not, as shown by eq. (12).

Another intriguing fact is that the interference term P(1∩2) given by eq. (10) can take up all real values including negative ones. This somehow jeopardizes our attempt to establish a correspondence with the third axiom of Kolomogorov. It is at this point that the classical axiom of positive probabilities is to be extended to include negative values as well. This leads to the extended probabilities. Similarly, the fact that the last two terms of eq. (9) are complex leads one to exotic or complex probabilities if we want to treat both these as probabilities on par with the first two terms. This is the proof of the richness of the possible extensions of traditional classical probabilities offered by the formalism in keeping with quantum theory.

However, the advantages of the formalism clearly far outweigh the disadvantages in the sense that it paves the way for a simple, logical and straightforward incorporation of the conscious subject in probability theory as well as in quantum theory. In our attempt to understand the Born rule from a deeper standpoint once we accept the duality axiom, everything falls in place and the objections based on classical notions of probability seem rather superficial. On the other hand, we may very well look ahead and relax the simplicity criterion by going for more complicated bilinear functional dependences of the probability on $A_s$ and $A_o$, while still respecting one or more of Kolomogorov's axioms. We may also make the assigning subject active rather than passive so that she can influence the objective amplitude $A_o$, and also the probability, by her subjective assignment $A_s$. These will certainly be issues of great importance and interest to scientists in both the fields in the future.

**Acknowledgements**

The author wishes to very sincerely acknowledge the fruitful discussions, especially on Wheeler's participatory universe paradigm, with Niranjan Barik and Lambodar P. Singh of Utkal University, Bhubaneswar and the technical support from the Institute of Physics, Bhubaneswar for carrying out the work.